\documentclass[conference]{IEEEtran}
\usepackage{amsmath,bm}
\usepackage{amsfonts}
\usepackage{mathrsfs}
\usepackage{amsbsy}
\usepackage{xcolor}
\usepackage{graphicx}
\usepackage{multirow}
\usepackage[ruled]{algorithm2e}

\DeclareMathOperator{\vect}{vec}

\ifCLASSINFOpdf
\else
\fi
\begin{document}

\title{Building Program Vector Representations\\ for Deep Learning}

%\author{\IEEEauthorblockN{Michael Shell}
%\IEEEauthorblockA{School of Electrical and\\Computer Engineering\\
%Georgia Institute of Technology\\
%Atlanta, Georgia 30332--0250\\
%Email: http://www.michaelshell.org/contact.html}
%\and
%\IEEEauthorblockN{Homer Simpson}
%\IEEEauthorblockA{Twentieth Century Fox\\
%\and
%\IEEEauthorblockN{James Kirk\\ and Montgomery Scott}
%\IEEEauthorblockA{Starfleet Academy\\
%San Francisco, California 96678-2391\\

% the acknowledgment of grants, issue a

\IEEEoverridecommandlockouts
\author{\IEEEauthorblockN{Lili Mou, Ge Li$^*$\thanks{$^*$Corresponding author.}, Yuxuan Liu, Hao Peng, Zhi Jin, Yan Xu, Lu Zhang
}
\IEEEauthorblockA{%\IEEEauthorrefmark{1}
Software Institute, School of EECS, Peking University\\
Beijing, 100871, P. R. China\\
Email: \{moull12, lige, zhijin, zhanglu\}@sei.pku.edu.cn\\
\{liuyuxuan, penghao.pku, alandroxu\}@gmail.com
}
}

\maketitle

\begin{abstract}
Deep learning has made significant breakthroughs in various fields of artificial intelligence.
Advantages of deep learning include the ability to capture highly complicated features,
weak involvement of human engineering, etc.
However, it is still virtually impossible to use deep learning to analyze programs
since deep architectures cannot be trained effectively with pure back propagation.
In this pioneering paper, we propose the ``coding criterion''
to build program vector representations, which are the premise of deep learning for program analysis.
Our representation learning approach directly makes deep learning a reality in this new field.
We evaluate the learned vector representations both qualitatively and quantitatively.
We conclude, based on the experiments,
the coding criterion is successful in building program representations.
To evaluate whether deep learning is beneficial for program analysis,
we feed the representations to deep neural networks, and achieve higher accuracy in the program classification
task than ``shallow'' methods, such as logistic regression and the support vector machine.
This result confirms the feasibility of deep learning to analyze programs.
It also gives primary evidence of its success in this new field.
We believe deep learning
will become an outstanding technique for program analysis in the near future.
\end{abstract}



% \fi
%\IEEEpeerreviewmaketitle

\section{Introduction}\label{sIntroduction}

Machine learning-based program analysis has been studied long in
the literature \cite{ml2,ml,dynamicAnalysis}.
Hindle et al. compare programming languages to natural languages and
conclude that programs also have rich statistical properties \cite{naturalness}.
These properties are difficult for human to capture,
but they justify using learning-based approaches to analyze programs.

The deep neural network, also known as \textit{deep learning},
has become one of the prevailing machine learning approaches since 2006 \cite{fast}.
It has made significant breakthroughs in a variety of fields,
such as natural language processing \cite{scratch,RNN},
image processing \cite{imagenet,image2}, speech recognition \cite{speech,speech2}, etc.
Compared with traditional machine learning approaches,
deep learning has the following major advantages:
\begin{itemize}
\item The deep architecture can capture highly complicated (non-linear) features efficiently.
    They are crucial to most real-world applications.
\item Very little human engineering and prior knowledge is required.
Interestingly, with even a unified model,
 deep learning achieves better performance than state-of-the-art approaches in many heterogeneous tasks \cite{unified}.
\end{itemize}
Such striking results raise the interest of its applications in the field of
program analysis.
Using deep learning to automatically capture program features
is an interesting and prospective research area.

Unfortunately, it has been practically infeasible for deep learning to analyze programs up till now.
Since no proper ``pretraining'' method is proposed for programs,
deep neural networks  cannot be trained effectively
with pure back propagation \cite{laywise,difficulty,strategy}
because gradients would either vanish or blow up through the deep architecture \cite{rnndifficult}.
No useful features can be extracted, and it results in very poor performance.

In this paper, we propose a novel ``coding criterion''
to build program vector representations based on abstract syntax trees (ASTs).
The vector representations are the premise of deep architectures, and
our method directly makes deep learning a reality in the new field---program analysis.
In such vector representations, each node in ASTs (e.g. \verb'ID', \verb'Constant')
is mapped to a real-valued vector, with each element
indicating a certain feature of the node.
The vector representations serve as a ``pretraining'' method.
They can emerge, through a deep architecture,
high-level abstract features, and thus benefit ultimate tasks.
We analyze the learned representations both qualitatively and quantitatively. We conclude from the experiments that the coding criterion is successful
in building program vector representations.

To evaluate whether deep learning can be used to analyze programs,
we feed the learned representations to a deep neural network in the program classification task.
We achieve higher accuracy than ``shallow'' methods.
The result confirms the feasibility of neural program analysis.
It also sheds some light on the future of this new area.

We publish all of our source codes, datasets, and learned representations on our project website\footnote{
https://sites.google.com/site/learnrepresent/
} to promote future studies.
The AST node representations can be used for further researches in
various applications of program analysis.
The source codes contain a versatile infrastructure of the feed-forward neural network,
based on which one can build one's own deep neural architectures.

To our best knowledge, this paper is the first to propose representation learning algorithms
for programs. It is also the first to analyze programs by deep learning.
This study is a pioneering research in the new field.
To sum up, the main contributions of this paper include:
\begin{enumerate}
\item Introducing the techniques of deep learning and representation learning
to the field of program analysis;
\item Proposing the novel ``coding criterion'' to build program representations, which
are the premise of deep learning;
\item Exploring the feasibility to analyze programs by deep neural networks,
shedding some light on the future;
\item Publishing all of our source codes, datasets, and learned representations online to promote further researches.
\end{enumerate}

In the rest of this paper, we first motivate our research in Section \ref{sMotivation}.
The background of deep learning and representation learning is introduced in Section \ref{sBackground}.
Then we explain our approach in detail in Section \ref{sModel}.
Experimental results are shown in Section \ref{sExperiment}.
In Section \ref{sFuture}, we look forward to the future of deep learning in the field of program analysis.
Last, we draw our conclusion in Section \ref{sConclusion}.

\bigskip
\section{Motivation}\label{sMotivation}

\subsection{From Machine Learning to Deep Learning}
Traditional machine learning approaches largely depend on human feature engineering,
e.g., \cite{bug} for bug detection, \cite{sim} for clone detection.
Such feature engineering is label-consuming
and \textit{ad hoc} to a specific task.
Further, evidence in the literature suggests
that human-engineered features may be even worse than automatically learned ones.
Mnih et al. report---for example, in the field of natural language processing (NLP)---the
automatically learned taxonomy of words \cite{hierarchical2} is better in their application than the famous WordNet
constructed by experts \cite{wordnet} used in \cite{hierarchical}.

Such results pose increasing demands on highly automated learning approaches,
such as deep learning with very little human engineering.

With deep neural networks,
program analysis may be easier with statistical methods.
For example, in the task of program classification,
deep neural networks automatically extract program features of interest.
Features can be organized hierarchically, from local to abstract.
Based on these abstract features,
we may determine the category of a program.
Such deep learning architectures require less human engineering
than existing methods like \cite{classification}.
Moreover, the feature representation nature also makes it easy
for multi-task learning.
As pointed out in \cite{MLAPP},
``many decision problems can be reduced to classification.''
Such deep learning architecture is also applicable to other program analysis tasks,
including
\begin{itemize}
\item bug detection as \cite{bug}, to which the deep learning approach is to
automatically extract features of bugs;
\item clone detection as \cite{sim2}, to automatically match the features of two programs;
\item code retrieval as \cite{sim4}, to automatically match program features with that of the queries; and
\item code recommendation as \cite{recsys}, to automatically predict the probability of the next possible codes, e.g. APIs, according to previous ones based on the features
    (like \cite{LM} in NLP).% and
%\item code categorization as \cite{cat}, to automatically cluster the features and thus codes.
\end{itemize}

In short, deep neural networks are capable of capturing highly complicated features
with little human involvement.
Analyzing programs with deep learning is an interesting and promising
research topic.

\subsection{Barriers of Deep Learning for Program Analysis}

Although deep neural networks are powerful enough to capture complicated features,
there are still barriers to overcome before they can be practically used to analyze programs.

Since all program symbols (e.g. nodes in ASTs) are ``discrete,''
no order is defined on these symbols.
Such discrete symbols cannot be fed directly to a neural network.
A possible solution is to map each symbol to a real-valued vector
in some dimension.
Each element in the vector characterizes a certain feature of the symbol spontaneously.
Hence, it is also referred to as \textit{distributed representation}.
By ``distributed,'' it is contrary to one-of-all coding, such as $k$-means clustering results.

A direct mapping approach is to randomly initialize these vector representations and train
with pure back propagation (like shallow networks).
Chances are that we end up with both poor optimization and poor generalization
if the network is deep \cite{laywise,difficulty,strategy}.
An alternative is to first learn the representations unsupervisedly
regardless of the specific task like clone detection, bug detection, etc.
Then they are fed to a neural network for supervised training.
The vector representations specify meaningful features of the symbols, and
benefit the ultimate tasks.
Hence, many researches focus on the problem of representation learning \textit{per se},
such as \cite{LM,energyLM,fast,RNN,improve} in fields like NLP.

However, due to the structural differences between natural languages and
programming languages \cite{PLNL},
existing representation learning algorithms in NLP are improper for programs.
As we know, natural languages are always written/spoken in one dimension as time flows.
By contrast, programmers always organize their source codes with proper indentation,
indicating branches, loops or even nested structures.
It will be extremely difficult to read source codes if they are written in one line (like natural languages). The formal grammar rules of the programming language
alias the notion of neighborhood. To be concrete, physically neighboring stuffs in a program source
code are not necessarily near to each other, but those in one grammar rule are.
%Therefore, we consider it impropriate to apply directly word representation learning algorithms
%to programs.

Further, if we want to build the representations for abstract components
of a program (e.g., function declaration/call nodes in ASTs),
existing NLP representation learning algorithms are inapplicable
since all of them are ``flat.''

Therefore, new approaches are needed to build program vector representations,
which are unfortunately not studied.

\bigskip
The above facts motivate our research of representation learning for programs.
This eventually makes deep learning a reality to analyze programs.
Considering current evidence in the literature,
we believe deep learning will
 make great progress in heterogenous tasks of program analysis.

\bigskip
\section{Background of Deep Learning and Representation Learning}\label{sBackground}

\subsection{Deep Neural Networks}
Nowadays, the
deep neural network is a widely-used technique in artificial intelligence.
Comprehensive reviews include \cite{AI,RL}.

\begin{figure}[!t]
\small
\centering
\includegraphics[width=.28\textwidth]{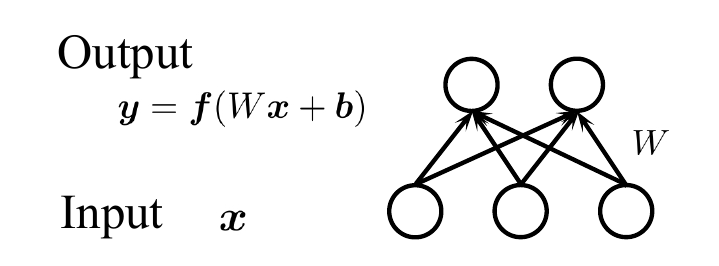}\\
(A) A single layer of neurons

\bigskip
\includegraphics[width=.47\textwidth]{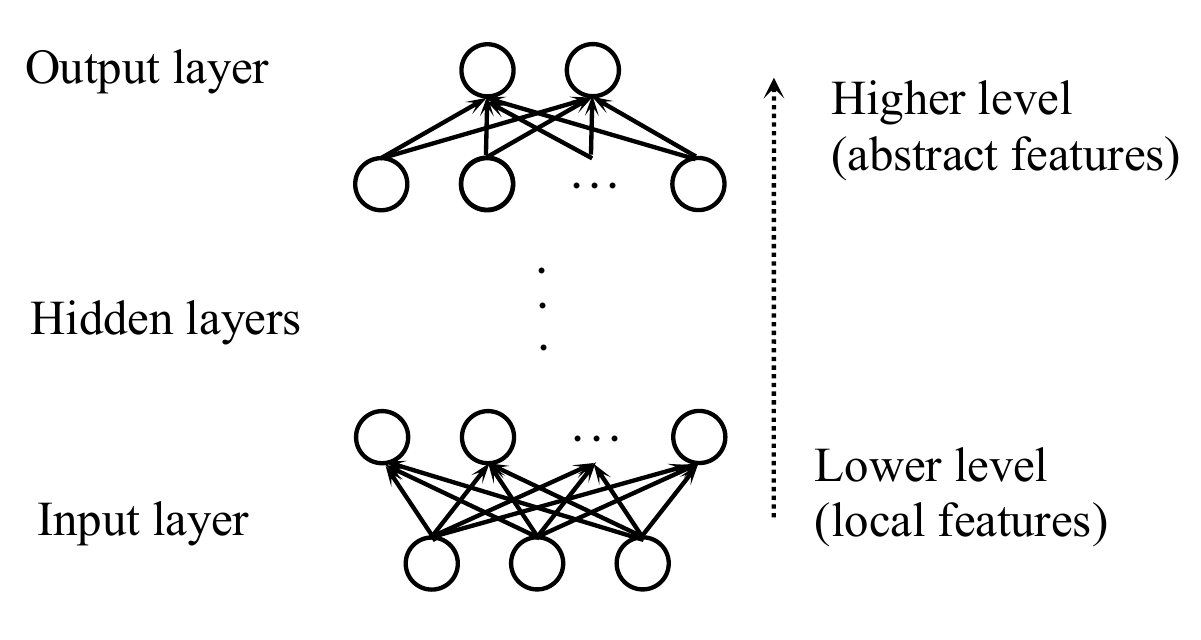}\\
(B) A deep neural network
\caption{A deep neural network (B) is composed of multiple layers of neurons (A).}
\label{fDL}
\end{figure}

A single layer of neurons, the building block for deep neural networks, takes a vector $\bm x \in \mathbb{R}^n$ as input and outputs a vector
$\bm y\in\mathbb{R}^m$ (Part A in Figure \ref{fDL}). Typically $\bm y$ is computed as

\begin{equation}
\label{eNeuron}
\bm y= \bm f(W\!\cdot\!\bm x+\bm b)
\end{equation}

\noindent where $W \in \mathbb{R}^{m\times n}, b\in\mathbb{R}^{m}$ are model parameters,
which are first randomly initialized and then trained by gradient descent, i.e., $W\leftarrow W-\alpha\frac{\partial J}{\partial W}$ and
$\bm b\leftarrow\bm b-\alpha\frac{\partial J}{\partial \bm b}$. ($\alpha$ is the learning rate; $J$ is the cost function.)
$\bm f$ is the activation function, usually non-linear, such as
$\operatorname{sigmoid}$, $\tanh$, etc.
The power of a single-layer neural network is limited: the decision boundary is linear,
and it is insufficient for most real-world applications.

Multi-layer neural networks (Part B in Figure \ref{fDL}) stack multiple layers of neurons.
The parameters can also be trained by gradient descent with back propagation algorithm \cite{machinelearning}.
Due to the stacking of multiple non-linear layers, multi-layer neural networks gain much more power.
It can be proved that a 2-layer\footnote{
The input layer is not counted to the number of layers in our terminology as
there is no weight associated with it.
} neural network with sufficient
hidden units can approximate arbitrary Boolean or continuous functions, and that
a 3-layer network can approximate any function \cite{approx}.
However, the shallow architecture is inefficient
because the number of hidden units may
grow exponentially in order to learn complicated (highly non-linear) features of data \cite{circuit}.
Such exponentiation of hidden layer units, and hence parameters, raises the VC-dimension of the model,
leading to very poor generalization \cite{learningTheory}.
%This phenomenon is also referred to as the \textit{curse of dimensionality} \cite{curse}.

The theory of circuits suggests deep architectures would be more efficient to capture complicated features \cite{AI}. In such deep architectures,
features can be organized hierarchically,
with local features at lower layers and
abstract features at higher layers (Figure \ref{fDL}).
However, while deep architectures capture abstract features efficiently,
they also make the networks very difficult to train. Few successful researches were reported
in early years using deep architectures (except convolutional neural networks \cite{lenet}).
%It was also suggested in text books (e.g., \cite{todo}) not stack more than two hidden layers
%for artificial neural network.

%

In 2006, Hinton et al. proposed stacked restricted Boltzmann machine (RBM) as a greedy layer-wise pretraining method for deep neural networks \cite{fast}.
During the pretraining phase, the stacked RBM is learning underlying data features unsupervisedly by minimizing
the energy function defined on the unlabeled data (i.e., maximizing the likelihood of the data).
Shortly after that, stacked autoencoders are used for pretraining \cite{laywise}, the criterion of which is to minimize the reconstruction error.
During the pretraining phase with either stacked RBMs or autoencoders,
the weights for neuron connections are learned, which extract underlying features of the data.
Then, for supervised learning, the neural weights are initialized
as that have been learned in the pretraining phase instead of random initialization.
Standard back propagation algorithm can be used for fine-tuning the weights
so as to be specific to the task.

The pretraining phase plays a vital role in deep learning.
It learns the features of data unsupervisedly, and as a result, the weights are much more meaningful than just chosen randomly.
According to the experiments reported in \cite{laywise,difficulty,strategy}, pretraining
helps optimization (minimizing the training error) as well as generalization (minimizing the test error).

However, the advantages of deep neural networks are not exploited in the field of program analysis.
We believe deep learning will also exhibit its power in this new field.
%Considering the evidences in the literature along with the primary evidence
%shown in Section \ref{sExperiment},
%we believe deep learning will also make great progress in
%heterogeneous tasks in program analysis, such as bug detection, clone detection, etc.

\subsection{Existing Representation Learning Approaches in NLP}
Neural networks and the pretraining approaches like RBMs, autoencoders
work well with image data, speech data, etc.
But they cannot be applied directly to the field like NLP and program analysis
because words and program symbols are ``discrete.''

In data like images, a total order is defined on the value.
For example, a gray-scale pixel with value 0 is black. If the value increases, it becomes brighter accordingly. If the value is 255, the pixel becomes white. However,
in fields like NLP, a word with index 20 is by no means twice as large as the word with index 10 for any meaningful feature.
Therefore, unlike traditional approaches in NLP, where each words is treated as an atomic unit,
it is meaningless to feed the indexes of words to neural networks.
(Note the multiplication $W\!\cdot\!\bm x$ in Equation \ref{eNeuron}.)

Real-valued vector representations come to our rescue.
With such representations,
each word is mapped to a $k$-dimensional vector ($k=30,100,300$, etc), representing
certain (anonymous) features of the word.
The value reflects the degree that a feature is satisfied.
These word representations can be fed forward to standard neural networks, and
every routine of deep learning works.

Early word representation learning is related to language modeling, the objective of which is to
maximize the joint probability of a linguistic corpus.
In \cite{LM}, they predict the probability of each word given $n-1$ previous words.
By maximizing the conditional probability of the $n$-th word, useful word features are learned.
Hinton et al. introduce 3 energy-based models in \cite{energyLM},
where they learn word representations by
minimizing the energy (maximizing the likelihood) defined on neighboring words.
In \cite{hierarchical,hierarchical2}, hierarchical architectures are proposed to reduce the computational cost in calculating the probabilities.
Later, researchers realized the normalization of probability is not essential
if we merely want to learn feature vectors.
Fast algorithms are then proposed in \cite{word2vec,word2vec2}.

All the above approaches adopt the Markovian assumption, where each word is related to
finite many previous words. Such approaches take into consideration
 only local patterns of physically nearing words.
Recurrent neural network (RNN) is introduced
in order to capture long-term dependencies \cite{recurrent}.
However, RNN may be very difficult to train
since the gradient would either vanish
or blow up during back propagation \cite{rnndifficult}.
Besides, the time-delaying nature of RNN treats data as a one-dimensional signal,
where structural information is also lost.

As we have discussed in Section \ref{sIntroduction},
programming languages are different from natural languages
in that the former contain richer and more explicit structural information.
Therefore, new representation learning algorithms are needed.
To solve this problem, we propose the ``coding criterion'' based on ASTs.
The detail of our approach is explained in the following section.

\bigskip
\section{The Coding Criterion for Program Representation Learning}\label{sModel}

In this section, we first discuss the granularities of program representation.
We settle for the granularity
of nodes in abstract syntax trees (ASTs).

In Subsection \ref{ssFormalize}, we formalize our approach and give the learning objective.
In Subsection \ref{ssLearn}, we present the stochastic gradient descent algorithm for training.

\subsection{The Granularity}\label{ssGranularity}

We should first answer a fundamental question of program representation learning---the
granularity of representation.
As we have introduced in previous sections,
vector representations map a symbol to a real-valued, distributed vector.
Possible granularities of the symbol include character-level, token-level, etc.
We analyze each case as follows.
\begin{itemize}
\item \textbf{Character-level.} Characters are the most atomic units of programming languages.
At this level, we treat each character in source codes as a symbol. This means that
we should learn the representations for characters like a-z, A-Z, 0-9 and all punctuation marks.
Although some researches explore character-level modeling for NLP \cite{characterlevel},
it is improper for programming languages.
In NLP, the knowledge of word morphology can be explored to some extent by character-level modeling,
but the situation changes in programs.
For example, the token \verb'double' in a C code refers to a data type.
But if one writes \verb'doubles', it is an identifier (e.g., a function name),
completely different from \verb'double'.
However, \verb'double' and \verb'doubles' share most characters,
which leads to similar vector representations.
\begin{figure}[!t]
\centering
\begin{verbatim}
       double doubles(double doublee){
           return 2 * doublee;
       }
\end{verbatim}

(A) A C code snippet

\bigskip
\includegraphics[width=.38\textwidth]{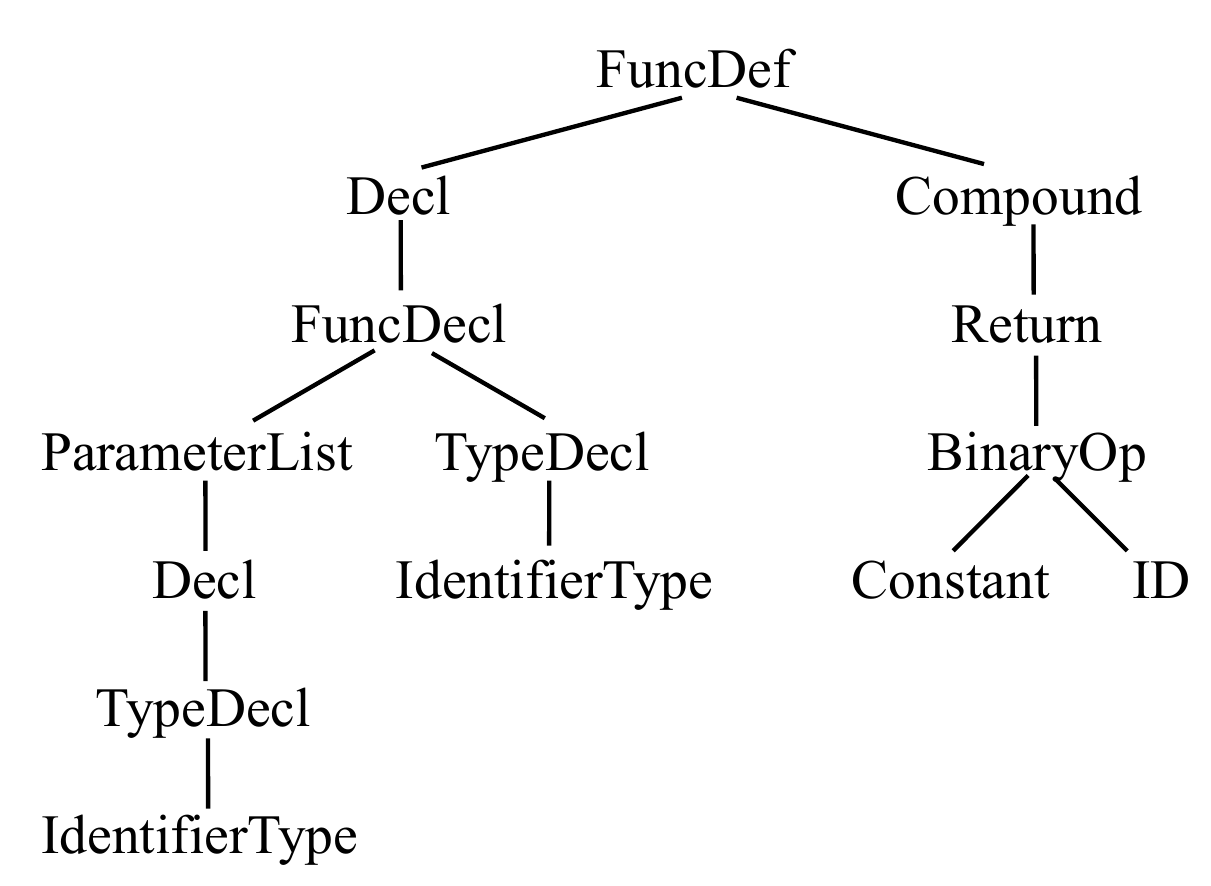}

(B) The corresponding AST
\caption{A C code snippet (A) and its corresponding AST (B).
Each node in AST corresponds to an abstract component (e.g., a function declaration,
a binary operator) in the program.
}\label{fAST}
\end{figure}

\item \textbf{Token-level.} This level is most similar to NLP representation learning.
We learn representations for all tokens (analogous to words in NLP), including types like \verb'int', \verb'double',
identifiers like \verb'doubles', \verb'func', etc.
Unfortunately, the identifier representations bring severe problems.
Unlike natural languages, where the number of words is generally fixed,
programmers can declare their own identifiers in their source codes, e.g., \verb'func1', \verb'func2', \verb'func3',
so on and so forth. Therefore, the number of tokens is unlimited.
Because some identifiers may appear only a few times (e.g., \verb'tmp'), we will suffer from
the problem of undesired data sparseness.
Hence, it is improper for representation learning at this level.

Another problem of token-level representation is that information is not encoded efficiently.
For example, we need two tokens to represent a pair of parentheses, indicating the
priority of different stuffs. In fact, such information need not to be expressed
explicitly in a more compressed representation like ASTs.

\item\textbf{Nodes in ASTs.} The abstract syntax tree is a structural representation of a program.
Figure \ref{fAST} shows a C code snippet and its corresponding AST, parsed by pycparser\footnote{
https://pypi.python.org/pypi/pycparser/
}.
At this level, we learn the representations for nodes in ASTs, e.g., \verb'FuncDef', \verb'ID', \verb'Constant'.
As is stated, the AST is more compressed compared with token-level representation.
Furthermore, there are only finite many types of nodes in ASTs, which makes it feasible to learn.
The tree structural nature of ASTs also provides opportunities to
capture structural information of programs.

Despite the above facts, the AST representation also has its drawback
since we regard all identifiers
as a same symbol.
Such codes as \verb'a*b' and \verb'c*d' cannot be distinguished between each other.
We hypothesize that structural information captures the semantics of programs to
a large extent.
For example, if we see two nested for-loops, inside of which is a branch of comparison followed
by three assignments, the code snippet is likely to be an implementation of bubble sort.
This level is also used in traditional
program analysis like code clone detection \cite{cc,ast}, vulnerability extrapolation \cite{ast2}, etc.
The experimental results in Section \ref{ssQuantitative} also suggest high accuracy in the program classification task at this level.

\item\textbf{Statement-level, function-level or higher.} Theoretically,
a statement, a function or even a program can also be mapped to a real-valued vector.
However, such representations cannot be trained directly.
A possible approach of modeling such complex stuff is by composition, i.e.,
the representation of a complex stuff is composited by that
of atomic ones. Such researches in NLP is often referred to as \textit{compositional semantics}
\cite{grounded}.
The state-of-the-art approaches in NLP compositional semantics can
only model sentences, paragraphs roughly.
It is very hard to capture the precise semantics; the ``semantic barrier''
is still not overcome.
\end{itemize}

To sum up, we have analyzed in this part different granularities of program representations.
We think the representation for nodes in ASTs has theoretical foundations,
and is feasible to learn and useful in applications.
In the following subsection, we formalize our
coding criterion to build program vector representations.

\subsection{Formalization}\label{ssFormalize}

%Such mapping should not be arbitrary.
The basic criterion of representation learning is that
similar symbols should have similar representations.
Further, symbols that are similar in some aspects should have similar values
in corresponding feature dimensions.
This is referred to as ``disentangling the underlying factors of variation''
in \cite{RL}.

In our scenario of representation learning for AST nodes,
similarity is defined based on the following intuition.
We think such symbols as \verb'ID',
\verb'Constant' are similar because both of them are related to data reference;
\verb'For', \verb'While' are similar because both are related to loops.
The observation is that similar symbols have similar usages in the programming language:
both \verb'ID' and \verb'Constant' can be an operand of a unay/binary operator;
 both \verb'For' and \verb'While' are a block of codes, etc.

To capture such similarity using AST structural information, we propose the ``coding criterion''.
The idea is that the representation of a node in ASTs should be ``coded''
by its children's representations via
a single neural layer.

We denote the vector of node $x$ as $\vect(x)$.
$\vect(\cdot)\in\mathbb{R}^{N_f}$, where $N_f$ is the dimension of features.
($N_f$ is set to 30 empirically in our experimental setting.)
For each non-leaf node $p$ in ASTs and its direct children
$c_1, \cdots, c_n$, their representations are $\vect(p), \vect(c_1), \cdots, \vect(c_n)$.
The primary objective is that

\begin{equation}
\label{eApprox}
\vect(p) \approx \tanh\left( \sum_{i=1}^n l_iW_i\cdot\vect(c_i) +\bm b\right)
\end{equation}

\noindent where $W_i\in \mathbb{R}^{N_f\times N_f}$ is the weight matrix for node $c_i$;
$\bm b\in\mathbb{R}^{N_f}$ is the bias term.
The weights ($W_i$'s) are weighted by the number of leaves under $c_i$ and the coefficients are

\begin{equation}
l_i = \dfrac{\#\text{leaves under } c_i}{\#\text{leaves under } p}
\end{equation}

As we have noticed in Figure \ref{fAST},
different nodes in ASTs may have different numbers of children,
and thus the number of $W_i$'s is hard to determine.
To solve this problem, one extreme is to apply
dynamic pooling \cite{dynamic,multilingual}.
In this method, we take the summed or maximal value over $\vect(c_i)$ for each feature dimension,
and thus only one weight matrix is needed.
This is also mathematically equivalent to the continuous bag-of-words model \cite{word2vec}.
However, when pooling is applied,
position information of $c$'s  will be totally lost and therefore it is not satisfactory.
Another extreme is to assign a different matrix parameter for each different position \cite{grounded}.
This method works in the original application with dependency trees, but may fail
in our scenario because there will be too many weights.

What we propose is a model called continuous binary tree, where there are two weight
matrices as parameters, namely $W_l$ and $W_r$.
Any weight $W_i$ is a linear combination of the two matrices.
That is, regardless the number of children, we treat it as a ``binary'' tree.
Formally, if $p$ has $n$ ($n\ge2$) children, then for child $c_i$,

\begin{equation}
W_i = \frac{n-i}{n-1}W_l + \frac{i-1}{n-1}W_r
\end{equation}

\noindent If $n=1$, $W_i=\frac{1}{2}W_l +\frac{1}{2}W_r$.

\begin{figure}[!t]
\centering
\includegraphics[width=.3\textwidth]{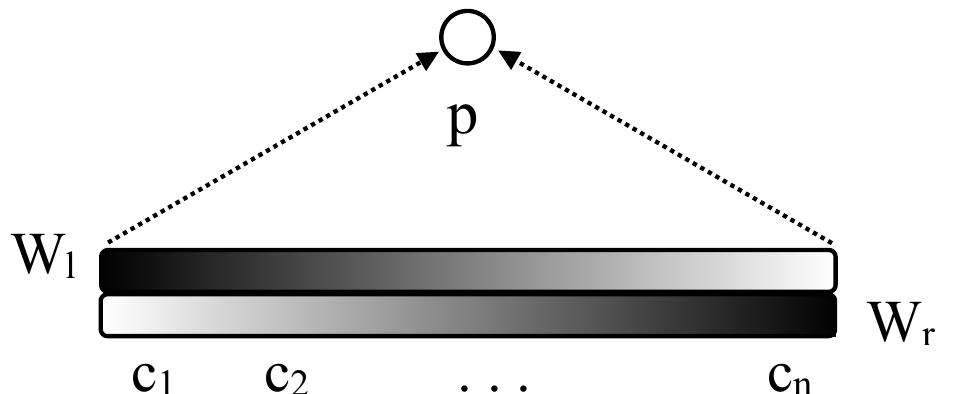}
\caption{Illustration of continuous binary tree. There are two weight matrices $W_l$ and $W_r$.
The gray-scale bars at the bottom indicate the coefficients of the weight parameters
($W_l$ and $W_r$ respectively) at the corresponding position.
}\label{fCBT}
\end{figure}

This process is illustrated in Figure \ref{fCBT}, where
the gray-scale values in the two bars represent the weight coefficients
for the node at the corresponding position.
With this model, the relative position information of a node can be coded into the network.

Now that we are able to calculate the weight $W_i$ for each node
and thus the right-hand side of Equation \ref{eApprox},
we measure closeness by the square of Euclidean distance, as below:

\begin{equation}
d = \left\|  \vect(p) - \tanh\left(\sum_{i=1}^n l_iW_i\cdot \vect(c_i)+
\bm b\right)\right\|_2^2\label{eD}
\end{equation}

According to our ``coding criterion,'' $d$ needs to be as small as possible.
However, we cannot directly minimize Equation \ref{eD}. Otherwise,
the network is likely to output trivial representations like $\vect(\cdot)=\bm 0, W=\bm 0, \bm b=\bm0.$
Such result gives zero distance but is meaningless.

To solve the problem, negative sampling can be applied \cite{unified,grounded,reasoning}.
The idea is that for each data sample $x$, a new negative sample $x_c$ is generated.
Since $x_c$ violates the patterns of valid data,
it needs to have a
larger distance (denoted as $d_c$) than $d$.
Hence, negative sampling method is also sometimes referred to as the pairwise ranking criterion \cite{ranking}.
In our program representation learning, we randomly select a symbol (one of $p$, $c_1$, $c_2$, $\cdots$, $c_n$)
in each training sample
and substitute it with a different random symbol.
The objective is that $d_c$ should be at least as large as $d+\Delta$,
where $\Delta$ is the margin and often set to 1.
The error function of training sample $x^{(i)}$ and its negative sample $x_c^{(i)}$ is then

\begin{equation}
J(d^{(i)},d_c^{(i)}) =  \max\left\{0, \Delta+d^{(i)}-d_c^{(i)}\right\}
\end{equation}

To prevent our model from over-fitting,
we can add $\ell_2$ regularization to weights ($W_l$ and $W_r$).
The overall training objective is then

\begin{table*}[!t]
\centering
\normalsize
\caption{Examples of the nearest neighbor query results.}
\label{tNN}
\begin{tabular}{|c|lcr|}
\hline
\multirow{2}{*}{\textbf{$\!$Query$\!$}}& \multicolumn{3}{c|}{\textbf{Results}}\\
\cline{2-4}
$\!$&$\!$\textbf{Most Similar}$\!$&$\!$$\!$&$\!$\textbf{Most Dissimilar}$\!$$\!$\\
\hline
$\!$ID$\!$&$\!$BinaryOp, Constant, ArrayRef, Assignment, StructRef$\!\!$&$\!\!\!\cdots\!\!\!$&$\!\!$PtrDecl, Compound, Root, Decl, TypeDecl$\!$\\
\hline
$\!$Constant$\!$&$\!$ID, UnaryOp, StructRef, ArrayRef, Cast$\!\!$&$\!\!\!\cdots\!\!\!$&$\!\!$EnumeratorList, ExprList, If, FuncDef, Compound$\!$\\
\hline
$\!$BinaryOp$\!$&$\!$ArrayRef, Assignment, StructRef, UnaryOp, ID$\!\!$&$\!\!\!\cdots\!\!\!$&$\!\!$PtrDecl,
Compound, FuncDecl, Decl, TypeDecl$\!$\\
\hline
$\!$ArrayRef$\!$&$\!$BinaryOp, StructRef, UnaryOp, Assignment, Return$\!\!$&$\!\!\!\cdots\!\!\!$&$\!\!$Compound, PtrDecl, FuncDecl, Decl, TypeDecl$\!$\\
\hline
$\!$If$\!$&$\!$For, Compound, Break, While, Case$\!\!$&$\!\!\!\cdots\!\!\!$&$\!\!$BinaryOp, TypeDecl, Constant, Decl, ID$\!$\\
\hline
$\!$For$\!$&$\!$If, While, Case, Break, Struct$\!\!$&$\!\!\!\cdots\!\!\!$&$\!\!$BinaryOp, Constant, ID, TypeDecl, Decl$\!$\\
\hline
$\!$Break$\!$&$\!$While, Case, Continue, Switch, InitList$\!\!$&$\!\!\!\cdots\!\!\!$&$\!\!$BinaryOp, Constant, TypeDecl, Decl, ID$\!$\\
\hline
$\!$While$\!$&$\!$Switch , Continue , Label , Goto$\!\!$&$\!\!\!\cdots\!\!\!$&$\!\!$
BinaryOp, Constant, Decl, TypeDecl, ID$\!$\\
\hline
$\!$FuncDecl$\!$&$\!$ArrayDecl, PtrDecl, FuncDef, Typename, Root$\!\!$&$\!\!\!\cdots\!\!\!$&$\!\!$ArrayRef, FuncCall, IdentifierType, BinaryOp, ID$\!$\\
\hline
$\!$ArrayDecl$\!$&$\!$FuncDecl, PtrDecl, Typename, FuncDef, While$\!\!$&$\!\!\!\cdots\!\!\!$&$\!\!$BinaryOp, Constant, FuncCall, IdentifierType, ID$\!$\\
\hline
PtrDecl$\!$&$\!$FuncDecl, Typename, FuncDef, ArrayDecl$\!\!$&$\!\!\!\cdots\!\!\!$&$\!\!$FuncCall, ArrayRef, Constant, BinaryOp, ID$\!$\\
\hline
\end{tabular}
\end{table*}

\begin{align}
\nonumber
\operatorname*{minimize}\limits_{W_l, W_r, \bm b}\;\; &
\dfrac{1}{2N}\sum\limits_{i=1}^N J(d^{(i)},d_c^{(i)})\\
\label{eObjective}
+&\dfrac{\lambda}{2M}
\Big(\parallel\! W_l\!\parallel_F^2 + \parallel\! W_r\!\parallel_F^2\Big)
\end{align}

\noindent where $N$ is the number of training samples; $M=2N_f^2$ is the number of weights
(number of elements in $W_l$ and $W_r$);
$\parallel\cdot\parallel_F$ refers to Frobenius norm;
$\lambda$ is the hyperparameter that strikes the balance between coding error and $\ell_2$ penalty.

\subsection{Training}\label{ssLearn}
The numerical optimization algorithm we use is stochastic gradient descent with momentum.
The model parameters $\Theta=\Big(\vect(\cdot), W_l, W_r, \bm b\Big)$ are first initialized randomly.
Then, for each data sample $x^{(i)}$ and its negative sample $x_c^{(i)}$,
we compute the cost function according to Formula \ref{eObjective}.
Back propagation algorithm is then applied to compute the partial derivatives and
the parameters are updated accordingly.
This process is looped until convergence.
The coding criterion of vector representation learning---as a pretraining phase for neural program analysis---is ``shallow,''
through which error can back propagate. Thus, useful features are learned for AST nodes.

To speed up training, we adopt the momentum method, where
the partial derivatives of the last iteration is added to the current ones with decay $\epsilon$.
Algorithm \ref{aGradient} shows the pseudo-code of the training process.

\begin{algorithm}[!t]
\caption{StochasticGradientDescentWithMomentum}\label{aGradient}
\KwIn{Data samples $x^{(i)}$, $i=1..N$;\\
\ \ \ \ \ \ \ \ \ Momentum $\epsilon$;\\
\ \ \ \ \ \ \ \ \  Learning rate $\alpha$}
\KwOut{Model parameters $\Theta=\Big(\vect(\cdot), W_l, W_r, \bm b\Big)$ }
Randomly initialize $\Theta$\;
\While{not converged}{
    \For{ i = 1..N }{
        Generate a negative sample $x_c^{(i)}$ for $x^{(i)}$\;
        Propagate forward and backward to compute $J^{(i)}$ and the partial derivative
        $\dfrac{\partial J^{(i)}}{\partial \Theta}$\;

        \medskip
        $\dfrac{\partial J^{(i)}}{\partial \Theta} \leftarrow \epsilon\dfrac{\partial J^{(i-1)}}{\partial \Theta}+\dfrac{\partial J^{(i)}}{\partial \Theta}$;\ \ \ \ // momentum

        \medskip
        $\Theta \leftarrow \Theta-\alpha\dfrac{\partial J^{(i)}}{\partial \Theta}$;
         \ \ \ \ \ \ \ \ \ \ \ \ \ \ \ // gradient descent

        \medskip
    }
}
\end{algorithm}

\bigskip
\section{Experiments}\label{sExperiment}
We first evaluate our learned representations by nearest neighbor querying and $k$-means clustering.
These qualitative evaluations give an intuitive idea about our vector representations.
We then perform supervised learning in the program classification task.
We feed the learned representations forward to deep neural networks.
The experimental results show that meaningful representations, as a means of pretraining,
make the network much easier to train in deep architectures.
We also achieve higher accuracy with the deep, tree-based convolutional neural network
compared with baseline methods.
We consider this as primary evidence of the success of deep learning for program analysis.

The dataset, source codes and learned representations can be downloaded at our
project website.

\subsection{Qualitative Evaluation: Nearest Neighbor Queries and $k$-means Clustering}

As we have stated in Section \ref{ssFormalize}, similar nodes in ASTs (like \verb'ID', \verb'Constant')
should have similar representations.
To evaluate whether our coding criterion for representation learning has accomplished this goal,
we perform nearest neighbor queries.

For each query of a symbol, we sort all other symbols by distance (measured in Euclidean space).
Examples are presented in Table \ref{tNN}.
As we can see, \verb'ID' and \verb'Constant' are the nearest neighbor of each other.
This seems meaningful since both of them are related to data reference. Similar symbols also include \verb'ArrayRef',
\verb'BinaryOp', which
are related to data manipulating.
Symbols like \verb'If', \verb'For', \verb'While', \verb'Break' are similar
as they are related to control flow. \verb'FuncDecl', \verb'ArrayDecl', \verb'PtrDecl'
are similar as they are declarations.
Moreover, these three groups are dissimilar with each other. (See most dissimilar part
in Table \ref{tNN}.)

To further confirm the above potential clusters with vector representations, we perform $k$-means clustering, where $k$ is set to 3. The result is shown in Table \ref{tKmeans}.
As we see, almost all the symbols in Cluster 1 are related to data reference/manipulating.
Cluster 2 is mainly about declarations. Cluster 3 contains more symbols, the majority
of which are related to control flow.
This result confirms our conjecture that similar symbols can be clustered into groups
with the distributed vector representations that are learned by our approach.

%To visualize the relationship between different symbols,
%we reduce the dimensionality of the vector representations by the t-SNE manifold learning algorithm
%\cite{}
%and plot the symbols in Figure \ref{todo}.

As the qualitative evaluations show, the learned representations
are meaningful as they can characterize the relationships between different symbols effectively.
The results are consistent with human understanding of programs.

It should also be reminded that similarity is not the only goal of representation learning.
Even though heuristic metrics can also be used to measure similarity---like \cite{esa} in NLP and \cite{sim,sim2} in program analysis,
which may be useful for
code clone detection \cite{sim3}, code retrieval \cite{sim4}---they
fail to capture different aspects of the relationships between different
symbols because the similarity is the only outcome of these metrics.
Thus, they are not suitable for highly-automated learning algorithms,
e.g., deep neural networks.
On the contrary, real-valued representations are distributed.
As each dimension captures a feature in a certain aspect spontaneously,
the distributed vector representations can emerge high-level abstract features and
benefit various tasks.
Therefore, representation learning
is crucial to program analysis with deep learning approaches.

\begin{table}[!t]
\caption{The result of $k$-means clustering. $k$ is set to 3.}
\label{tKmeans}
\centering
\normalsize
\begin{tabular}{|c|c|}
\hline
\textbf{$\!\!$Cluster$\!\!$} & \textbf{Sybmols}\\
\hline
\multirow{2}{*}{1} & UnaryOp, FuncCall, Assignment, ExprList,\\
&StructRef, BinaryOp, ID, Constant, ArrayRef\\
\hline
\multirow{2}{*}{2} & FuncDef, TypeDecl, FuncDecl, Compound,\\
&  ArrayDecl, PtrDecl, Decl, Root\\
\hline
\multirow{7}{*}{3} & Typedef, Struct, For, Union, CompoundLiteral,\\
& TernaryOp, Label, InitList, IdentifierType,\\
& Return, Enum, Break, DoWhile, Case,\\
& DeclList, Default, While, Continue,\\
& ParamList, Enumerator, Typename, Goto,\\
& Cast, Switch, EmptyStatement,\\
& EnumeratorList, If\\
\hline
\end{tabular}
\end{table}

\subsection{Quantitative Evaluation: Improvement for Supervised Learning}\label{ssQuantitative}

We now evaluate whether building program vector representations is beneficial for
real-world tasks, i.e., whether they will improve optimization and/or generalization
for supervised learning of interest.
We feed the representations to
the Tree-based Convolutional Neural Network (TCNN) for program classification.

The dataset comes from an online Open Judge (OJ) system\footnote{
http://programming.grids.cn/
}, which contains a large number of programming problems for students.
Students solve the problems and submit their source codes to the system.
The OJ system automatically compiles, runs and judges the validity of the source codes.
We select four problems for our program classification task.
Source codes (in C programming language) of the four problems are downloaded along with their labels (problem IDs).
We split the dataset by $3:1:1$ for training, cross-validating (CV) and testing.

\begin{figure}[!t]
\centering
\normalsize
\includegraphics[width=.5\textwidth]{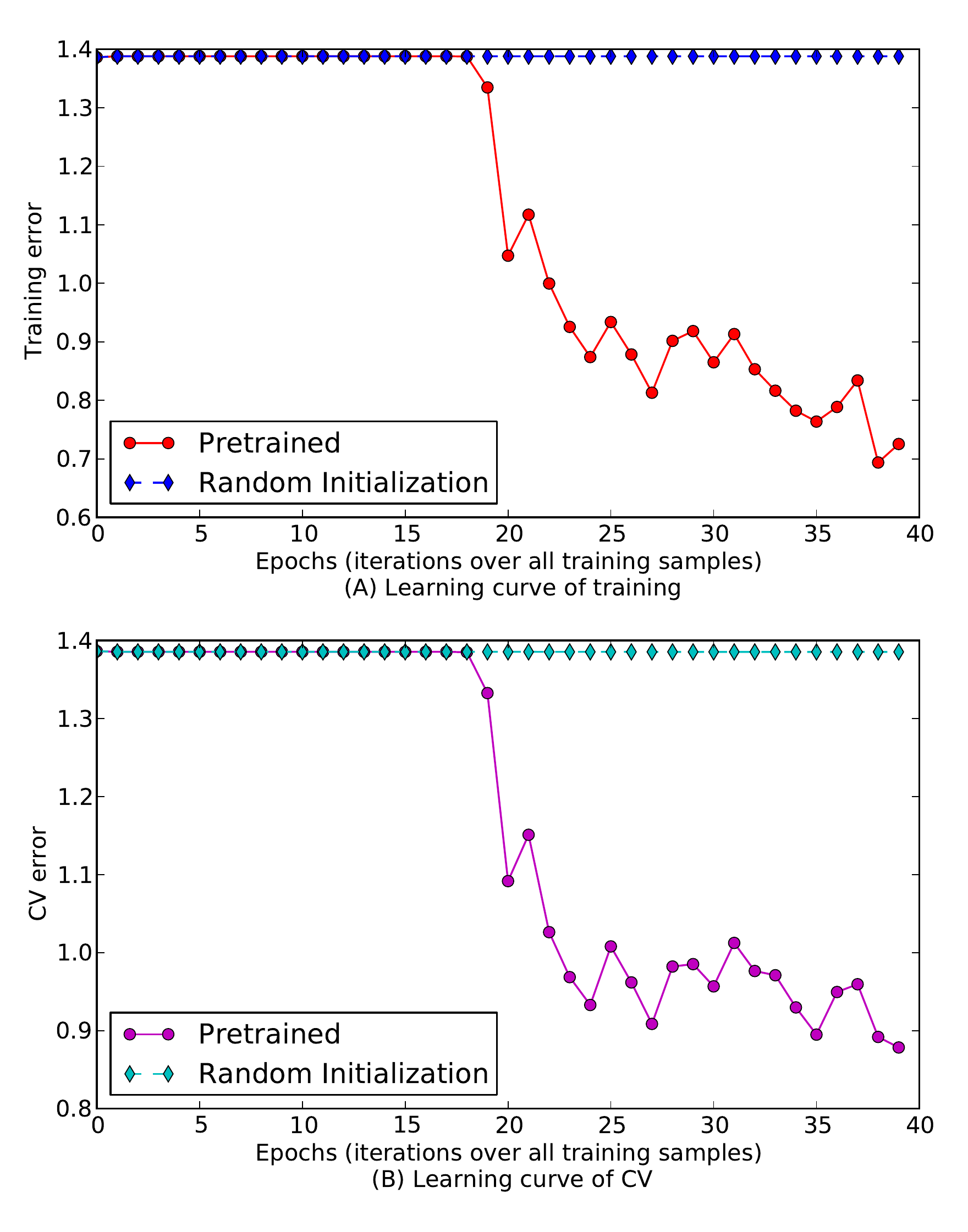}
\caption{Learning curves of training (A) and CV (B).
The learned program vector representations improve supervised learning in terms of both generalization and optimization.}\label{tSupervise}
\label{fPretrain}
\end{figure}

Figure \ref{fPretrain} plots the learning curves for training and CV
in first 40 epochs.
(One epoch is an iteration over all training samples.)
The X axis is the number of epochs during training.
The Y axis is the cross-entropy error, computed as

\begin{equation}
J = - \dfrac1N\sum_{i=1}^{N} \sum_{j=1}^M t_j^{(i)}\log y_j^{(i)}
\end{equation}

\noindent where $N$ is the number of data samples (training or CV respectively); $M=4$ is the number of labels
(different programming problems); $y_j$ is the probability for label $j$ estimated by
the TCNN model;
$\bm t$ is the actual label (one-of-all coding),
with $t_j$ indicating whether data sample $i$ belongs to label $j$.

Since no effective program representation existed before,
the deep TCNN model could not be trained at all, as the
blue curve demonstrates at the top of Part A in Figure \ref{fPretrain}.
(Here, all model parameters are initialized randomly,
which is a prevalent setting in ``shallow'' architectures.)
The reason is that gradients will 
vanish or blow up during back propagation through a deep network.
No useful features are learned, and as a result, TCNN also performs poorly on the CV set,
indicated by the cyan curve at the top of Part B in Figure \ref{fPretrain}.

On the contrary, the program representation serves as a pretraining method.
If the vector representations and the coding parameters,
namely $\vect(\cdot)$, $W_l$, $W_r$ and $\bm b$, are initialized as are learned by our
coding criterion,
the training and CV errors
decrease drastically (the red and magenta curves) after a plateaux of about 15 epochs, which leads
to the high performance of TCNN.

The fact that unsupervised pretraining improves supervised learning is also reported
in \cite{laywise,difficulty,strategy}, where RBMs and autoencoders are used
as pretraining methods for generic data (mainly the MNIST handwritten digit dataset in these papers).
As pretraining explores underlying data features unsupervised,
it gives a much more meaningful initialization of parameters.
Therefore, the deep neural networks can be trained much faster and more effectively.
Our experimental results in program analysis 
are consistent with these reports in the literature in other fields.

To evaluate whether deep learning may be helpful for program analysis, we compare TCNN to
baseline methods in the program classification task.
In these baseline methods,
 we adopt the bag-of-words model, which is a widely-used approach in text classification \cite{textmining}.
As shown in Table \ref{tSupervise}, logistic regression, as a linear classifier, achieves 81.16\% accuracy.
The support vector machine (SVM) with radial basis function (RBF) kernel explores non-linearity, and improves the result by 10\%.
By automatically exploring the underlying features and patterns of programs,
TCNN further improves the accuracy by more than $4\%$.
This experiment suggests the promising future of deep leaning approaches in the field of program analysis.

\begin{table}[!t]
\normalsize
\centering
\caption{Accuracy of Program Classification.}
\begin{tabular}{cc}
\hline
\hline
\textbf{Method} & \textbf{Accuracy}\\
\hline
Random guess & 25.00\%\\
Logistic regression & 81.16\%\\
SVM with RBF kernel & 91.14\%\\
TDNN (a deep learning approach) & \textbf{95.33}\%\\
\hline
\hline
\end{tabular}
\end{table}

\bigskip
To sum up, we evaluate the learned representations
empirically by nearest neighbor querying and $k$-means clustering.
%We also plot the symbols in 2-D to visualize the result.
Program classification experiment shows
the learned representations are greatly beneficial for supervised learning.

Based on the above experiments,
we conclude that the proposed ``coding criterion'' based on ASTs
is a successful representation learning algorithm for programs.

Our experimental result in program classification
confirms the feasibility of using deep learning to analyze programs.
It also shows primary evidence of its success in the new field.

\bigskip
\section{Looking Forward to the Future}\label{sFuture}

As evidence in the literature show,
deep learning is making breakthroughs in many fields of artificial intelligence.
We believe it will also become an important method in various tasks
in the field of program analysis.
As a pioneering study,
we address the following promising research topics in this new area.

\subsection{Different Perspectives for Program Modeling}
In this paper, we treat a program as a tree, where
each node corresponds to an ``abstract'' component of the program.
We hypothesize in this paper that structural information is important
to programs to a large extent, and our experiments confirm our conjecture.
However, the AST is not the only perspective of program modeling.

Another possible perspective is treating a program
as a sequence of statements.
Such perspective is also adopted in traditional program analysis, e.g.,
API usage pattern mining \cite{apimining,apimining2}.
As the representations can be composited by atomic symbols (e.g., AST nodes),
we can also apply deep learning approaches to sequences of statements.
Although some structural information may be lost,
the neighboring information are captured
and local patterns can be extracted.

Treating a program as a 2-dimensional signal is an interesting, novel and also meaningful perspective,
which is bionics-inspired.
As we, human beings, always read source codes on a 2-D screen,
it is also possible for neural networks to model programs in this perspective.
Indents and linefeeds on the 2-D screen are useful features because they
suggest strong semantics of programs.
Existing techniques in computer vision can be applied,
e.g. the convolutional neural network (CNN).
CNN is analogous to visual cortex of human brains, and thus
it has the solid biological foundation in cognitive science.
Interestingly, as a bionics-inspired model,
deep CNN achieved unexpected high performance \cite{lenet} before pretraining methods were invented.

\subsection{Integrating Prior about Programs to Network Architectures}

Despite the fact that a unified architecture of deep neural networks is applicable to various tasks
with high performance, we can also integrate human priors to the networks.

CNN is an example that specifies explicitly the physically neighborhood information of an image.
Physically neighboring pixels form local patterns (e.g. a circle, a line), which can be detected
by convolution kernels. Being fed forward to higher layers in the network,
the local patterns emerge high-level abstract features.
Such abstract features are beneficial for the ultimate task (e.g., object recognition).
Another widely-used domain specific prior in deep learning is slowness \cite{slow,slow2}.
As it is not desired that features of image/acoustic data are changing too fast,
penalties of variation are added to the cost function, so that
the learned features are ``slow.''

For program analysis, priors can also be integrated to the neural networks.
In one of our undergoing research, we would like to capture the local features of ASTs.
A tree-based convolutional neural network (TCNN) is proposed and studied.
Primary results have been reported in Section \ref{ssQuantitative}.

Another prior is that we can integrate formal methods of program analysis
to neural networks (or vise versa). \cite{knowledgebase} is an example of neural reasoning for knowledge base.
For programs, even though all non-trivial properties are undecidable,
formal methods can be viewed as an approximation with pure mathematical deduction, often giving the guarantee of either no false-positive, or no false-negative, which may be important to program analysis \cite{formal}.
For now, it seems hard to combine these two techniques,
but once they were combined, it would be beneficial for both.

\subsection{Various Applications}

The application of deep learning in the field of program analysis is another
promising research topic,
which is at least as important as, if not more important than, the theory of deep learning.
Some are pointed out in Section \ref{sIntroduction}, including
code clone detection, bug detection, and code retrieval.
%Due to the main goal of this paper, the details are not to be discussed here.

\bigskip
In short, as deep learning is brand new to program analysis,
the questions addressed in this part still remain unknown to the literature.
It is not clear which perspective is most proper to model programs,
or which is most suitable for what application.
It is also not very clear how to integrate human priors about programs to
the neural network architecture.
These are among the fundamental questions of deep learning when it is applied to the new field.
Besides, real-world applications of deep learning are also important for program analysis.

\bigskip
\section{Conclusion}\label{sConclusion}

In this paper, we study deep learning and representation learning
in the field of program analysis.
We propose a novel ``coding criterion'' to build vector representations of nodes in ASTs,
which make deep learning a reality for program analysis.
We also feed the learned representations to a deep neural network to classify programs.

The experimental results show that our representations successfully
capture the similarity and relationships among different nodes in ASTs.
The learned representations significantly improve supervised training
for deep neural networks
in terms of both optimization and generalization.
We conclude that the coding criterion is successful in building program vector representations.
The experiments also confirm the feasibility of deep learning to analyze programs,
and show primary evidence of its success in the new field.

As a pioneering study, we address several fundamental problems,
including the perspectives of program modeling, the integration of human priors
and the applications of deep learning.

To promote further researches in the new field,
we publish all of our datasets, source codes, and learned representations online.

We believe, considering the fact that deep learning has made breakthroughs in many fields of
artificial intelligence, along with the primary evidence reported in this paper,
deep learning will become an outstanding approach of program analysis
in the near future. We call for more studies in this new, prospective field.

\newpage




\bibliographystyle{IEEEtran}
\bibliography{dl,se}

\end{document}